\documentclass[showpacs,aps,prl,twocolumn]{revtex4-1}
\usepackage{graphicx} 
\usepackage{dcolumn}
\usepackage{bm}
\usepackage[mathscr]{euscript}
\usepackage{amssymb,amsmath}
\usepackage{epstopdf}
\usepackage{color}
\usepackage[english]{babel}
\usepackage{ulem}

\begin{document}

\title{Parity violation in ferromagnet-superconductor heterostructures \\
with strong spin-orbit coupling}


\author{P. M. Svetlichnyy, Z. Jiang, and C. A. R. S\'a de Melo}
\affiliation{School of Physics, 
Georgia Institute of Technology, 
Atlanta, GA 30332, USA}

\date{\today}

\begin{abstract}
We study spectroscopic properties of ferromagnetic-superconductor 
heterostructures with strong spin-orbit coupling of the Rashba type and in 
the presence of exchange fields. The superconducting layer (film) experiences 
both an intrinsic spin-orbit field and an exchange field due to the 
proximity to ferromagnetic layers (films). We analyse the temperature 
dependence of the order parameter for superconductivity at various values of 
exchange field and spin-orbit coupling, and describe momentum-dependent 
properties that exhibit parity violation. Furthermore, we show that parity 
violation can be probed in tunneling experiments of the single-particle 
density of states and in photoemission experiments of the 
momentum distribution. 
\end{abstract}


%
%

\maketitle

%
%

The interplay of magnetism and superconductivity has played a very 
important role in several materials including 
Cuprate~\cite{dagotto-1994} and Pnictides~\cite{johnston-2010}, 
while the interplay between spin-orbit effects and 
superconductivity has been important in the case of 
non-centro-symmetric~\cite{sigrist-2012} and topological~\cite{zhang-2011} 
superconductors. In the case of Cuprate and Pnictides the interplay 
of magnetism and superconductivity leads to a very rich phase diagram 
and to unconventional behavior such as d-wave and multi-s-wave order 
parameters, but there is no evidence that the superconducting ground 
states of these systems violate parity. Similarly in the case of 
non-centro-symmetric or topological superconductors, where 
spin-orbit coupling (SOC) terms lead to parity-odd matrix elements, 
ground state properties do not exhibit parity violation.

%
%

\begin{figure}[t]
\centering
\includegraphics[width=0.45\textwidth]{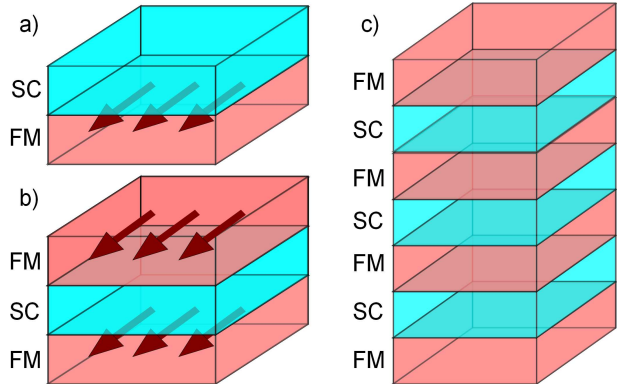}
\caption{
(color online) Layered heterostructures consisting 
of ferromagnets (FM) and superconductors (SC) with spin-orbit coupling
in (a) bilayer, (b) trilayer, and (c) multilayer configurations.
}
\label{fig:one}
\end{figure}

In this paper, we study a simple case of the interplay of magnetism, 
superconductivity and spin-orbit coupling to a superconducting ground 
state that violates parity. 
In Fig.~\ref{fig:one}, we show possible geometries for 
the realization of such ground state. We focus on the simpler bilayer 
and trilayer cases shown in Figs.~\ref{fig:one}(a),
and ~\ref{fig:one}(b), respectively. However,
the effect also can exist in the multilayered systems illustrated in 
Fig.~\ref{fig:one}(c).
In the bilayer and trilayer cases, we show that when the 
spin-orbit-coupled superconducting layer experiences a strong 
in-plane exchange field due to the proximity to a 
ferromagnetic layer, it may no longer exhibit properties with well 
defined parity. Parity is violated in the superconducting 
layer when its critical temperature is lower than the ferromagnetic 
ordering temperature.
This parity violation  manifests itself in spectral properties of 
the superconducting layers such as the quasi-particle excitation spectrum, 
single-particle density of states, and spin-dependent momentum distribution. 
The latter two properties may be measured via tunnelling and 
photoemission experiments. Our theoretical findings point into a new
experimental direction for ferromagnet-superconductor 
multilayers, beyond the traditional proximity effects~\cite{buzdin-2005},
magnetic couplings across the superconducting layer~\cite{sademelo-1997},
and Josephson coupling across a ferromagnetic layer~\cite{kontos-2002}.
Our results are particularly relevant to recent experimental results
that show the emergence of exchange interactions in 
ferromagnet-superconductor multilayers consisting of Manganites and 
Cuprates~\cite{santamaria-2012a,santamaria-2012b}, where if SOC exists, 
then parity violation will also emerge.

%
%

%
%

\begin{figure}[b]
\includegraphics[width=0.45\textwidth]{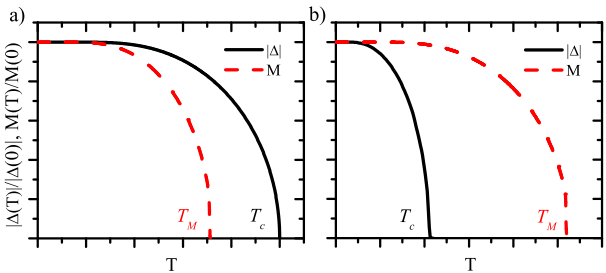}
\caption{
(color online) 
Sketches of  normalized dependencies 
of the magnetization $M (T)/M(0)$ (red-dashed line) 
and the order parameter 
$\vert \Delta (T) \vert/ \vert \Delta (0) \vert$ 
(black-solid line) as functions of temperature T. 
Two cases are possible: 
(a) $T_{M} < T_{c}$, (b) $T_{M} > T_{c} $.
}
\label{fig:two}
\end{figure}

In the heterostructures (multilayered systems) shown in 
Fig.~\ref{fig:one}, the layers that become 
superconducting (SC) experience a transition 
from their normal (N) to their superconducting state at critical 
temperature $T_c$, while the layers that become ferromagnetic (FM) 
experience a transition from a paramagnetic (PM) to a ferromagnetic 
state at the Curie temperature $T_M$. 
In general, two cases are possible, both represented in 
Fig.~\ref{fig:two}. In case I, shown in 
Fig.~\ref{fig:two}(a), the Curie temperature $T_{M}$ 
is lower than critical temperature $T_{c}$ of the superconductor,
that is, $T_{M} < T_{c}$. In this case, the Curie temperature is low, 
which is unfrequently found among existing materials.  
In case II, shown in Fig.~\ref{fig:two}(b), the order of
temperatures is $T_{M} > T_{c}$, which is more likely to be 
experimentally relevant in the immediate future, and, thus, 
we focused our specific calculations to this case.

%
%

For the bilayer and trilayer cases the Hamiltonian density of the 
combined ferromagnet-superconductor (FM-SC) system has three contributions 
$
{\cal H} ({\bf r}) 
= 
{\cal H}_S ({\bf r}) + {\cal H}_{FS} ({\bf r}) 
+ {\cal H}_F ({\bf r}).
$
The first term of ${\cal H} ({\bf r})$ is 
\begin{equation}
{\cal H}_S ({\bf r}) = 
\sum_{s,s^\prime}
\psi^{\dagger}_{s}({\bf r})
[\hat{K}{\bf 1} + {\bf H}_{SO}(-i\nabla)]_{ss'}
\psi_{s^\prime}(\mathbf{r})
+ {\cal H}_I ({\bf r})
\end{equation}
describing the superconducting layer, 
with $\hat{K} = -\nabla^{2}/2m - \mu$ being the kinetic energy, 
$[{\bf H}_{SO}(-i\nabla)]_{ss'}$ being the SOC, 
and 
$
{\cal H}_I 
= 
- g\psi^{\dagger}_{\uparrow}({\bf r})\psi^{\dagger}_{\downarrow}
({\bf r})\psi_{\uparrow}({\bf r})\psi_{\downarrow}({\bf r})
$
being the local interaction term. 
Here, $\psi^{\dagger}_{s}({\bf r})$ is a creation operator of an 
electron with spin $s$ located at the point ${\bf r}$. 
The second term 
$
{\cal H}_{FS} ({\bf r}) 
=  
- J_{FS,\nu}\psi^{\dagger}_{s}({\bf r}) 
[\sigma_{\nu}]_{ss^\prime} 
\psi_{s^\prime} ({\bf r})
\cdot S_{F,\nu} ({\bf r} + {\bf a}),
$
where $S_{F,\nu}$ is a spin in the ferromagnetic layer 
located at distance ${\bf a}$ away from the superconducting layer 
along the c-axis of the heterostructure, $\sigma_{\nu}$ is the Pauli
matrix, and $J_{FS, \nu}$ is the exchange coupling along the $\nu$ direction.
The third term 
$
{\cal H}_F ({\bf r}) 
= 
K_F ({\bf r}) 
- 
J_{\mu \nu} 
\sum_{i \ne j}
S_{F,\mu} ({\bf r}_i)
\cdot 
S_{F,\nu} ({\bf r}_j)  
$
describes the ferromagnetic layers, which can be itinerant (localized) 
if the magnetic state is metallic (insulating) with non-zero (zero)
kinetic energy density $K_F ({\bf r})$. In either case, when 
in-plane ferromagnetism sets in at $T_M$, 
the SC layers experience a strong parallel exchange
field. For definitess, we assume that the FM layer is insulating
and governed by a magnetic Hamiltonian density corresponding to 
the $XYZ$ model:
$
{\cal H}_F ({\bf r})
= 
- J_\nu \sum_{<ij>} 
{\bf S}_{F,\nu} ({\bf r}_i)
\cdot 
{\bf S}_{F,\nu} ({\bf r}_j), 
$
where $\nu = x, y, z$. For simplicity, we take the case of 
$J_y \gg \{J_x, J_z\}$, such the magnetization points essentially
along the $y$-direction of the superconductor~\cite{nota-bene-1}.

In this case, the effective Hamiltonian matrix for the superconducting layer 
acquires the simple form
\begin{equation}
\label{eqn:hamiltonian-matrix}
{\bf H}_0({\bf k})
=
\left(
\begin{array}{cccc}
\widetilde{K}_{\uparrow}({\bf k}) & -h^{*}_{\perp}({\bf k}) & 0  & -\Delta \\
-h_{\perp}({\bf k}) & \widetilde{K}_{\downarrow}({\bf k}) & \Delta & 0 \\
0 & \Delta^{*}& -\widetilde{K}_{\uparrow}(-{\bf k})& h_{\perp}({\bf -k})\\
-\Delta^{*} & 0 & h^{*}_{\perp}({\bf -k}) & -\widetilde{K}_{\uparrow}({\bf -k})  
\end{array}
\right),
\end{equation}
in the four-dimensional Nambu basis
$
\Psi^\dagger ({\bf k})
=
\left(
\psi^\dagger_\uparrow({\bf k}),
\psi^\dagger_\downarrow({\bf k}),
\psi_\uparrow({\bf -k}),
\psi_\downarrow({\bf -k})
\right).
$
Here, the kinetic energy for the $\uparrow$ spin is 
$
\widetilde{K}_{\uparrow}
= 
k^2/2m - \mu - h_z
$
and for the $\downarrow$ spin is
$
\widetilde{K}_{\downarrow}
= 
k^2/2m - \mu + h_z,
$
while the order parameter for superconductivity
$
\Delta 
=
\vert \Delta \vert e^{i\varphi}
$ 
with $\vert \Delta \vert$ being its magnitude, and with
$\varphi$ being its phase. The spin-flip field
$
h_{\perp}({\bf k})
=
h_{x}({\bf k})
+
ih_{y}({\bf k})
$
is the complex representation of the sum of the  components of the exchange 
and spin-orbit fields felt by the electrons in the superconducting layer.
The exchange fields are $h_\nu = J_\nu \langle S_{F ,\nu}\rangle$, where
$\nu = x, y, z$, while the spin-orbit fields are assumed to be of the Rashba
type~\cite{Rashba-1960} 
$
h_{R}({\bf k}) 
=
- v_{R}k_{y}
+ iv_{R}k_{x}.
$ 
Since $J_y \gg \{ J_x , J_z \}$ the magnetization in the FM layer points
along the $y$ direction, then $h_z = h_x = 0$, but $h_y \ne 0$, which leads
to the total spin-flip field
$
h_{\perp}({\bf k}) 
=
-v_{R}k_{y}
+i( h_y + v_{R}k_{x} ).
$

For simplicity, we consider only case II, in the limit of $T_M \gg T_c$, 
which is sufficient to produce a parity violating superconducting state.
In this case, the thermodynamic potential corresponding to 
${\bf H}_0 ({\bf k})$ defined in Eq.~(\ref{eqn:hamiltonian-matrix}) is
\begin{equation}
\Omega_0 
=
V\frac{\left|\Delta\right|^2}{g} 
-\frac{T}{2}
\sum_{{\bf k},j}
\ln
\left[
1 + \exp\left(-\frac{E_{j}({\bf k})}{T}\right)
\right]
+\sum_{{\bf k}}\widetilde{K}_{+}({\bf k}),
\end{equation}
where 
$
\widetilde{K}_{+}({\bf k})
=
\frac{1}{2}\left(\widetilde{K}_{\uparrow}({\bf k})
+
\widetilde{K}_{\downarrow}({\bf k})\right)
$
is a reference kinetic energy, and $E_j ({\bf k})$ 
are the eigenvalues of ${\bf H}_0 ({\bf k})$.
The saddle-point order parameter equation 
\begin{equation}
\label{eqn:order-parameter}
\frac{V}{g} \vert \Delta \vert
=
-\frac{1}{4}
\sum_{{\bf k},j} 
n_F \left( E_{j}({\bf k}) \right)
\frac{\partial E_{j}({\bf k})}{\partial \vert \Delta \vert}
\end{equation}
is obtained from the condition 
$
\delta \Omega_0
/
\delta \vert\Delta\vert 
= 0
$
while the number equation 
\begin{equation}
\label{eqn:number}
N_{0}
=
\sum_{{\bf k}}
\left(
1
-
\frac{1}{2}
\sum_{j} 
n_F \left( E_{j}({\bf k}) \right)
\frac{\partial E_{j}({\bf k})}{\partial\mu}\right)
\end{equation}
fixes the chemical potential $\mu$ and is obtained from the
relation 
$
N_{0} 
= 
-\partial\Omega_0/\partial\mu.
$
In the expressions above $n_F$ is the Fermi function.
These two equations need to be solved numerically and 
self-consistently~\cite{nota-bene-2}. 
In calculations we assume that the constants
have values
$
g/V
=
4.49 \times 10^{-6}\varepsilon_F
$
and
$
N_0/V
=
k_F^2/(2\pi)
$. 
The solution obtained is checked for the minimum condition 
$
\partial^{2}\Omega_{0}
/
\partial \vert \Delta \vert^{2} 
> 
0
$
to guarantee the thermodynamic stability of the system.

In order to solve for $\vert \Delta \vert$ and $\mu$,
we need to calculate explicitly the eigenvalues $E_j ({\bf k})$.
Notice that when there is no magnetization in the ferromagnet, 
the effective exchange fields are zero and all the matrix elements of 
${\bf H}_0 ({\bf k})$ have well-defined parity, i.e., 
$h_\perp ({\bf k})$ is odd, while
all the other elements are even in momentum space. However, when
the in-plane exchange field $(h_x,h_y, 0)$ is non-zero, the spin-flip
matrix element $h_\perp ({\bf k})$ does not have well defined parity, 
while all the other elements remain parity even. This fact alone, 
leads to the  emergence of a parity violating quasi-particle (quasi-hole) 
energy spectrum when there is an in-plane component of the total exchange
field. If the exchange field is zero (no ferromagnetism), or if the 
magnetization (exchange field) is only along the $z$-direction, then
there is no parity violation in the excitation spectrum and in the 
superconducting state. The solution for the excitation spectra can 
be obtained analytically, but the expressions are extremely cumbersome
in the parity violating case, thus we prefer to obtain the eigenvalues
of ${\bf H}_0 ({\bf k})$ numerically, by making a base transformation
$
\Phi ({\bf k})
=
{\bf U} ({\bf k})
\Psi ({\bf k})
$
where $\Phi ({\bf k})$ is the four-dimensional spinor 
describing quasi-particles and quasi-holes, and 
${\bf U} ({\bf k})$ is the unitary matrix of 
coherence factors that diagonalizes ${\bf H}_0 ({\bf k})$. 

%
\begin{figure}[b]
\includegraphics[width=0.45\textwidth]{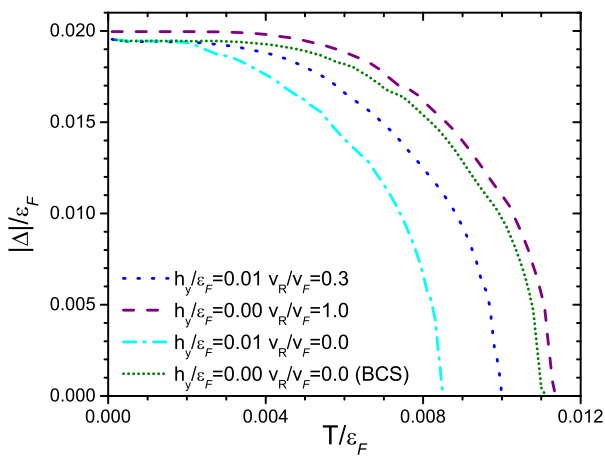}
\caption{(color online) 
Order parameter $\vert \Delta (T) \vert/\varepsilon_F$ versus
$T/\varepsilon_F$ for different 
values of the $T = 0$ magnetization in the ferromagnet 
(exchange field $h_y/\varepsilon_F$) and spin-orbit coupling 
constant $v_R/v_F$. 
For $T = 0$ zero CM momentum Cooper pairs become unstable at
$h_y/\varepsilon_{F}  > 0.02$, where pairing 
with non-zero CM momentum occurs.
} 
\label{fig:three}
\end{figure}
%

%
%

Numerical results for $\vert \Delta (T) \vert/\varepsilon_F$ 
are shown in Fig.~\ref{fig:three} for various values of $h_y/\varepsilon_F$ 
and $v_R/v_F$, where $\varepsilon_F$ $(v_F)$ is the Fermi energy (velocity) 
of the non-interacting Fermi gas without SOC. Notice that
increasing $h_y$ tends to suppress superconductivity with zero
center-of-mass (CM) momentum pairing due to pair breaking. 
However, finite SOC tends
to stabilize superconductivity since its momentum-dependent spin-flip field
induces a triplet component in the order parameter which counters the pair 
breaking effect. In order to see the violation of parity in the SC layers,
it is essential to measure momentum dependent quantities via spectroscopic
techniques. Thus, next, we investigate three spectroscopic quantities 
that contain valuable information about parity violation in the 
superconducting state. 

\begin{figure*}[t]
\includegraphics[width=0.95\textwidth]{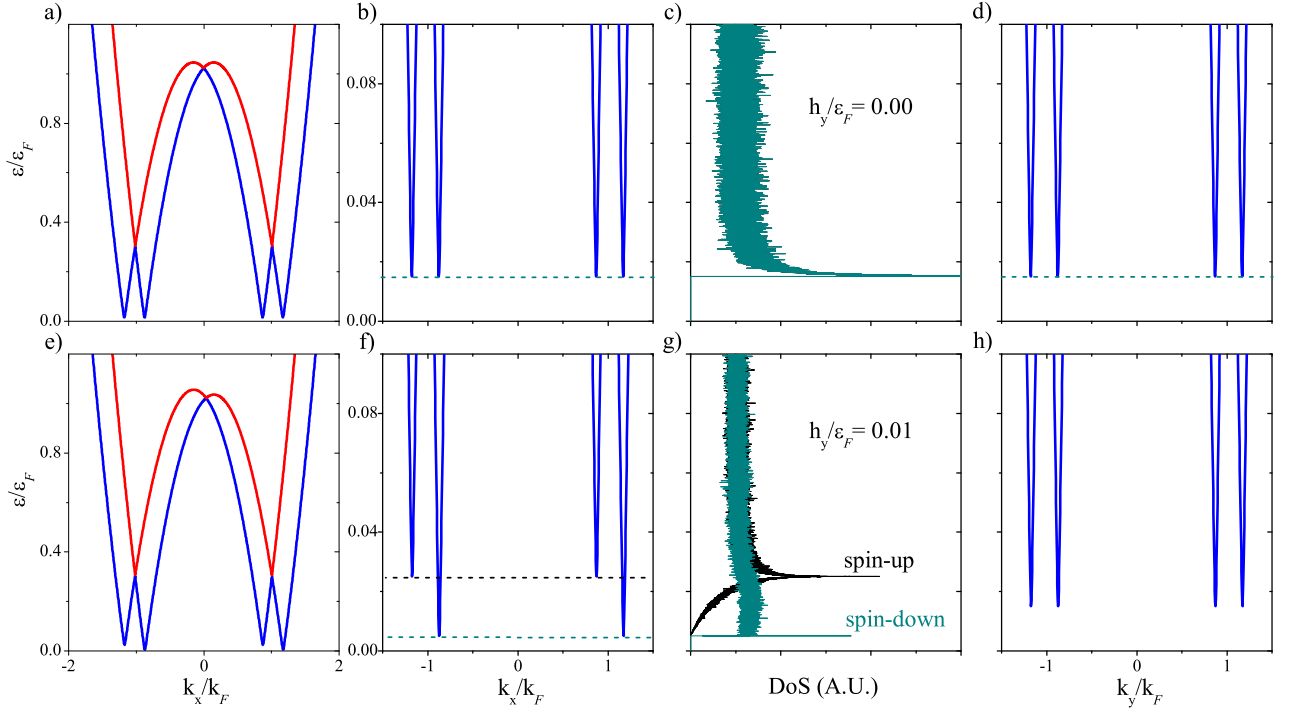}
\caption[width=0.7\textwidth]
{
(color online) 
Quasi-particle energy bands $E_j ({\bf k})$ 
along $\left(k_{x},0\right)$ are shown in (a), (e).
The low-energy behaviour along 
$\left(k_{x},0\right)$ and $\left(0,k_{y}\right)$ 
are shown in (b), (f) and (d), (h), respectively.
The corresponding spin-dependent density of states for electrons
are shown in (c) and (g). Note that these quantities differ 
from the quasi-particle density of states due to coherent
factors.  The set of parameters used is  
$T/\varepsilon_{F}=0.007$ $(T/T_c = 0.7)$, 
$ v_{R}/v_{F}=0.30$, $\mu/\varepsilon_{F}=1.02$, 
$\vert \Delta \vert /\varepsilon_{F}=0.015$ [(a)-(h)], 
$h_{y}/\varepsilon_{F}=0.00$ [(a)-(d)], 
$h_{y}/\varepsilon_{F}=0.01$ [(e)-(h)]. }
\label{fig:four}
\end{figure*}

The first property is the quasi-particle 
excitation spectrum consisting of the two upper branches of  
eigenvalues $E_j ({\bf k})$, shown in
Fig.~\ref{fig:four}, where a clearly parity violating
excitation spectrum is present when the magnetization (exchange field $h_y$)
is non-zero. The direction that affects the overall parity of the
excitation spectrum is $k_x$ because the magnetization of the ferromagnet
is assumed to point out along the $y$-direction alone. This leads to the 
spin-flip field 
$
h_{\perp}({\bf k}) 
=
-v_{R}k_{y}
+i( h_y + v_{R}k_{x} ),
$
that depends on the combination $h_y + v_R k_x$, and, therefore, does not
have well defined parity. More generally, when the 
magnetization of the ferromagnet has components along the 
$x$ and $y$ directions then 
$
h_{\perp}({\bf k}) 
=
(h_x - v_{R}k_{y})
+i( h_y + v_{R}k_{x} ),
$
and the excitation spectrum violates parity both along the $x$ and 
$y$ directions.

The second property is the electronic density of states (DOS), which can 
be obtained from the resolvent operator matrix 
$
{\bf G} ( i\omega, {\bf k} )
=
\left[
i\omega - {\bf H}_0 ({\bf k})
\right]^{-1}
$
in terms of the imaginary part of the diagonal elements 
of ${\bf G}$ as
$
\rho_{i} (\omega)
= 
- 
{(1/\pi)} 
{\rm Im}
{\bf G}_{ii} (\omega + i\delta, {\bf k}), 
$
where $i = (\uparrow, \downarrow)$ labels the  
spins in the original basis, 
with the spin quantization axis chosen to be along the perpendicular
direction $(z)$ to the films. The resulting expression in terms 
of the matrix of coherence factors is simply
\begin{equation}
\label{eqn:density-of-states}
\rho_{\Uparrow,\Downarrow} (\omega)
= 
\sum_{j, {\bf k}} 
\frac{1}{2}
\vert
{\bf U}_{1j} ({\bf k})
\pm
i{\bf U}_{2j} ({\bf k})
\vert^2
\delta 
\left(
\omega - E_j ({\bf k})
\right),
\end{equation} 
where $\Uparrow$ $(\Downarrow)$ corresponds 
to the up (down) spin of the particle with respect 
to the direction of the exchange field $h_y$.
The spin-dependent DOS is illustrated in 
Figs.~\ref{fig:four}(c) and~\ref{fig:four}(g).
Notice that Eq.~(\ref{eqn:density-of-states}) 
differs from the quasi-particle DOS
$
\rho_{qp, j} (\omega)
= 
\sum_{{\bf k}} 
\delta 
\left( 
\omega - E_j ({\bf k})
\right)
$
due to the presence of the coherence factors ${\bf U}_{ij}$. 
The main effects of $h_y \ne 0$ is to create a 
parity-violating asymmetry in the low-energy quasi-particle 
bands and produce the split-peak structure in the single-particle 
DOS seen in Fig.~\ref{fig:four}(g) in comparison to the $h_y = 0$ 
case shown in Fig.~\ref{fig:four}(c). Furthermore, the spin-up and 
spin-down electronic DOS are very different from each other as
the coherence factors are highly sensitive to the presence of the
exchange field $h_y$. We show only the $\omega > 0$ region as the 
$\omega < 0$ region
can be obtained by the transformation: 
$
\rho_{\Uparrow} (\omega) 
= 
\rho_{\Downarrow} (-\omega)
$
and 
$
\rho_{\Downarrow} (\omega) 
= 
\rho_{\Uparrow} (-\omega).
$
In the presence of $h_y$ the system remains gapped, but the 
induced triplet component of the order parameter in the generalized
helicity basis acquires a $k_x$ component in addition to the
$k_x + i k_y$ contribution when $h_y = 0$. The former contribution 
is responsible for the linear density of states right above the 
quasiparticle gap. Either the spin-dependent or total DOS 
$
\rho_T (\omega) 
= 
\rho_{\Uparrow} (\omega) + 
\rho_{\Downarrow} (\omega) 
$
may be measured via tunneling 
experiments~\cite{ref-TDOS-01, ref-TDOS-02, ref-SDOS-01, ref-SDOS-02}.

%
%

%
%

\begin{figure}[ht]
\includegraphics[width=0.45\textwidth]{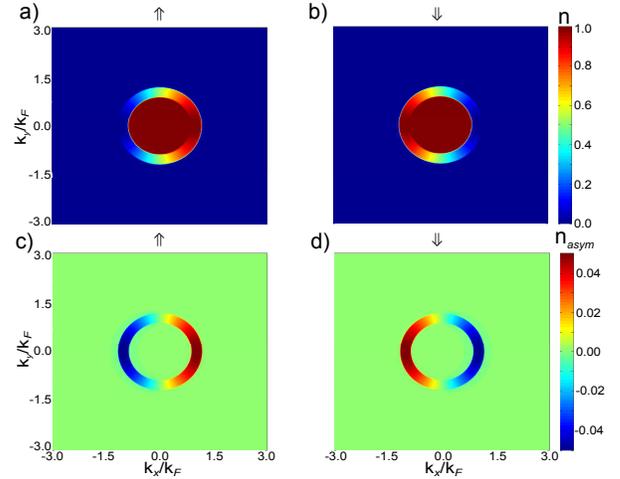} 
\caption{(color online) 
The momentum distributions $n_{\Uparrow} ({\bf k})$ and
$n_{\Downarrow} ({\bf k})$ are shown in (a) and (b), respectively.
Their corresponding asymmetric parts  
$
n_{i, asym} (\bf k)
$ 
are shown in (c) and (d).
The parameters used are: 
$T/\varepsilon_{F}=0.007$ $(T/T_c = 0.7)$, 
$v_{R}/\varepsilon_{F} =0.30 $, 
$h_{y}/\varepsilon_{F}=0.01$, 
$\mu/\varepsilon_{F}=1.02$, 
$\vert \Delta \vert/\varepsilon_{F}=0.015$.
}
\label{fig:five}
\end{figure}

The third property that we analyse is the spin-dependent
momentum distribution 
$
n_s({\bf k}) 
= 
\langle 
\psi_s^\dagger ({\bf k}) \psi_s ({\bf k}) 
\rangle,
$
where $s = (\Uparrow, \Downarrow).$
The resulting expression in terms of the coherence factors ${\bf U}_{ij}$
and the quasi-particle operators $(\phi^\dagger_j, \phi_j)$ is
$
n_{\Uparrow,\Downarrow}({\bf k})
=
\sum_{j}
\frac{1}{2}
\vert
{\bf U}_{1j} ({\bf k})
\pm
i{\bf U}_{2j} ({\bf k})
\vert^2
\langle 
\phi^{\dagger}_{j}({\bf k})
\phi_{j}({\bf k})
\rangle,
$
which can be further expressed in terms of the Fermi function 
$n_F$ and the eigenvalues $E_j ({\bf k})$ as
\begin{equation}
\label{eqn:momentum-distribution}
n_{\Uparrow,\Downarrow}({\bf k})
=
\sum_{j}
\frac{1}{2}
\vert 
U_{1j}({\bf k})
\pm
iU_{2j}({\bf k})
\vert^2
n_F 
\left( 
E_j ({\bf k})
\right).
\end{equation}
We show the spin-dependent momentum distributions in 
Fig.~\ref{fig:five} along with their asymmetric parts
$
n_{s,asym}
(k_{x},k_{y})
=
\frac{1}{2}
\left[
n_{s} (k_{x},k_{y})
-
n_{s} (-k_{x},k_{y})
\right].
$ 
The asymmetry of the distribution, which also represents parity violation, 
arises from the presence of both the in-plane 
exchange field and the in-plane Rashba SOC.
In the absence of either one of these terms, the 
momentum distribution would be parity even in both $k_x$ 
and $k_y$ directions. The total momentum distribution 
$ 
n({\bf k})
= 
n_\Uparrow ({\bf k}) 
+ 
n_\Downarrow ({\bf k})
$ 
is also parity violating, but the effect is smaller.  
Spin-dependent momentum distributions may be measured
via the recently developed spin- and angle-resolved 
photoemission spectroscopy (Spin-ARPES)
~\cite{ref-SARPES-00,ref-SARPES-01, ref-SARPES-02},
while total momentum distributions may be measured
via standard ARPES~\cite{ref-ARPES-01, ref-ARPES-02}.

%
%

In summary, we studied spectroscopic properties in ferromagnet-superconductor 
heterostructures and showed that a parity violating superconducting
state can exist, when the superconducting layers possess strong
spin-orbit coupling. We focused on the bilayered and trilayered
heterostructures, where the Curie temperature $T_M$ 
of the ferromagnetic layers was larger than the
critical temperature $T_c$ of the superconducting layers.
However, a similar effect also occurs when $T_M < T_c$, where
the superconducting state is first parity preserving, and then
below $T_M$ becomes parity violating. We found that 
if the in-plane Rashba spin-orbit coupling is zero, then the superconducting 
state is always parity preserving even if the ferromagnet is in its ordered 
state. Thus, we concluded that it is necessary to have both an in-plane 
magnetization (exchange field) and in-plane spin-orbit coupling 
for the emergence of a parity violating superconducting 
state.  Finally, we showed such parity violation can be detected through
the measurement of spectroscopic properties such as the quasi-particle 
excitation spectrum, single-particle density of states and spin-dependent 
momentum distributions. 
 
%
%

We thank NSF (DMR-0709584) for support.


\begin{thebibliography}{99}
\bibitem{nota-bene-1} This assumption is not necessary for the emergence of 
parity violation. Any in-plane magnetization will lead to parity violation
in the superconducting state.
\bibitem{nota-bene-2} Since $T_M \gg T_c$, the magnetization 
$M_\nu = \langle S_{F,\nu} \rangle$ in the FM layer is well 
developed and its major effect in the SC layer 
is to introduce a proximity induced exchange field with 
components $(h_x, h_y, h_z)$. The value of $M_\nu$ in the ferromagnet 
can be obtained separately. For instance, when 
$J_y \gg \{ J_x, J_z \}$, only $h_y \ne 0$, and the self-consistency 
equation for $h_y = J_y M_y$ is approximately 
%
$
h_y (T) 
= 
h_y (0)\
\tanh \left[ h_y(T)/T \right].
$
\end{thebibliography}
\end{document}